\documentclass[aps,prl,twocolumn,superscriptaddress,amssymb,floatfix] 
{revtex4}
\usepackage{graphicx}
\usepackage{color}
\usepackage{bm}

\begin{document}

\title{$0-\pi$ Transitions in a
Superconductor/Chiral magnet/Superconductor Junction}

\author{Thierry Champel}
\affiliation{
Institut f\"{u}r Theoretische Festk\"{o}rperphysik and DFG-Center for  
Functional Nanostructures,
   Universit\"{a}t Karlsruhe,
   D-76128 Karlsruhe, Germany }
\affiliation{Laboratoire de Physique et Mod\'{e}lisation des Milieux  
Condens\'{e}s, CNRS and Universit\'{e} Joseph Fourier, 25 Avenue des  
Martyrs, BP 166, F-38042 Grenoble, France}
\author{Tomas L\"{o}fwander}
\affiliation{
Institut f\"{u}r Theoretische Festk\"{o}rperphysik and DFG-Center for  
Functional Nanostructures,
   Universit\"{a}t Karlsruhe,
   D-76128 Karlsruhe, Germany }
\affiliation{Department of Microtechnology and Nanoscience - MC2,  
Chalmers University of Technology, S-41296 G\"{o}teborg, Sweden}
\author{Matthias Eschrig}
\affiliation{
Institut f\"{u}r Theoretische Festk\"{o}rperphysik and DFG-Center for  
Functional Nanostructures,
   Universit\"{a}t Karlsruhe,
   D-76128 Karlsruhe, Germany }
\affiliation{
Fachbereich Physik, Universit\"at Konstanz,
D-78457 Konstanz, Germany }

\date{June 22, 2007}

\begin{abstract}
  We study the $\pi $ phase in a
  superconductor-ferromagnet-superconductor Josephson junction, with a
  ferromagnet showing a cycloidal spiral spin
  modulation with in-plane propagation vector.  Our results reveal a
  high sensitivity of the junction to the spiral order and indicate
  the presence of 0-$\pi$ quantum phase transitions as function of the
  spiral wave vector.  We find that the chiral magnetic order
  introduces chiral superconducting triplet pairs that strongly
  influence the physics in such Josephson junctions, with potential
  applications in nanoelectronics and spintronics.
\end{abstract}

\maketitle

It is by now well established that an equilibrium superconducting
phase difference of $\pi$ can be arranged between two singlet
superconductors when separating them by a suitably chosen
ferromagnetic material \cite{Gol,Bar2007}. Transitions between the
$\pi$-state and the $0$-state of such S-F-S Josephson junctions have been
revealed in experiments through oscillations of the Josephson critical
current with varying thickness of the ferromagnet \cite{Kon2002} or
with varying temperature \cite{Rya2001}. The $\pi$ Josephson junction
is currently of considerable interest as an element complementary to
the usual Josephson junction in the development of functional
nanostructures \cite{baselmans99}, including superconducting
electronics \cite{ortlepp} and quantum computing \cite{ioffe}.

Recently, there has been a rapid progress in the field of chiral
magnetism \cite{bogdanov01,bode07,binz06,Thessieu97} that raises the expectations
for applications of chiral magnets in spintronics. Chiral
order occurs in inversion asymmetric magnetic materials
\cite{bode07,Thessieu97} that in the presence of spin-orbit coupling
give rise to a Dzyaloshinskii-Moriya interaction ${\bf D}_{ij} \cdot
({\bf S}_i\times {\bf S}_{j})$. This interaction favors a directional
non-collinear (spiral) spin structure of a specific chirality over the
usual collinear arrangement favored by the Heisenberg
exchange interaction $J_{ij} ({\bf S}_i\cdot {\bf S}_{j})$. A
well-studied \cite{binz06,Thessieu97} chiral magnet is the
transition-metal compound MnSi, with the spiral wave length $\Lambda
\approx 180$ \AA. Nanoscale magnets or magnetic systems with reduced
dimensionality that frequently lack inversion symmetry due to
interfaces and surfaces are expected to exhibit chiral magnetism \cite{bogdanov01}. This
has been confirmed by the recent observation \cite{bode07} of a spin
spiral structure (with $\Lambda \approx 12$ nm) in a single atomic
layer of manganese on a tungsten substrate.

\begin{figure}[b]
\centerline{
\includegraphics[width=0.85\columnwidth]{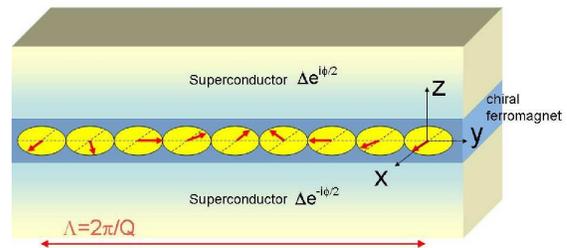}
}
\caption{(Color online)
S-CM-S Josephson junction where CM is a chiral ferromagnet with
a cycloidal spiral spin modulation, i.e. the spins are confined to a  
plane
(the $x-y$-plane) parallel to the spiral propagation direction
(the $y$-axis).  }
\label{Fig0}
\end{figure}

In this Letter, we combine chiral magnetism with superconductivity in
a controllable Josephson nano-device where $0-\pi$ transitions can be
induced by tuning the magnetic spiral wave vector $Q$ (see
Fig.~\ref{Fig0}). Whereas in bulk magnets $Q $
can be manipulated e.g. by means of pressure, in nano-magnets
alternative possibilities of control exist, as
electric fields, geometry, or pinning layers.
Such a Josephson device shows a surprisingly complex behavior with
$0$- to $\pi$-state transitions as function of spiral wave length
$\Lambda=2\pi/Q$, that turn into zero temperature transitions for some
critical wave vectors. However, below the threshold
$\Lambda_{\mathrm{th}} =\pi \xi_J$, where $\xi_J$ is the penetration
depth of pairs into the chiral magnet ($\xi_{J}$ depends on material
constants), a qualitatively different behavior is found.

Within our model chiral magnetism and singlet superconductivity take  
place in
mutually separated materials, and the magnetic spiral affects only the
superconducting proximity amplitudes. This is in contrast
to the case of coexisting superconducting and spiral magnetic order in
the same material, e.g. in ferromagnetic superconductors
\cite{kulic01}.  We also contrast our model to the case of a helical
spiral spin modulation with a propagation wave vector perpendicular to
the S-F interface \cite{Ber2001b},
and the Josephson effect in S-F-S junctions with a N\'{e}el domain
structure \cite{Fom2006}. The physics studied in
Refs. \cite{Ber2001b,Fom2006} is dominated by the presence of
long-range triplet components, that are absent in the present system
\cite{Cha2005}
(concerning the role of long-range triplet pairs in
S-F-S hybrid structures see also \cite{Esc2003}).
In Refs. \cite{Ber2001b,Fom2006}, a strong dependence of the Josephson
critical current $I_{c}$ on the ferromagnet inhomogeneity is found
due to these long-range components. However, the related magnitude of
$I_{c}$ is so small that the observation of such an
effect is questionable. In this paper, we report a critical current with
a much larger magnitude \cite{magnitudenote}, which is essential for
potential applications.

We study the S-CM-S junction shown in Fig.~\ref{Fig0} within the
framework of the quasiclassical theory of superconductivity and
consider the diffusive limit. Furthermore, we shall assume that the
pair correlations induced in the chiral magnet, quantified by the
anomalous Green function $f$, are small. This is fulfilled for
temperatures close to the superconducting critical temperature
$T_{c}$, and also for much smaller temperatures $T$ provided that the  
S-CM interface
transparency is small. We decompose the 2 $\times$ 2 spin-matrix
$f$ as $f=f_{s} i\sigma_{y} +i ({\bf f}_{t} \cdot {\bm
   \sigma}) \sigma_{y} $, where $f_{s}$ is the singlet component and
${\bf f}_{t}$ is the triplet vector (here ${\bm
   \sigma}=(\sigma_{x},\sigma_{y},\sigma_{z})$ is a vector of Pauli
matrices).  These components obey
\cite{Cha2005} in the magnet a system of linearized Usadel  
equations
\begin{eqnarray}
(D {\bm \nabla} ^{2} - 2 \varepsilon_{n}) f_{s} & =& 2 i {\bf J}  
\cdot {\bf f}_{t},  \label{S1}\\
(D {\bm \nabla} ^{2} - 2 \varepsilon_{n}) {\bf f}_{t}& =&2 i {\bf J}  
f_{s}, \label{S2}
\end{eqnarray}
where $\varepsilon_{n}=\pi T(2n+1)$ is the Matsubara frequency with
$n$ a positive integer. Quantities for negative frequencies are
obtained through symmetry relations \cite{Cha2005}, the components $f_ 
{s}$ and $ {\bf f}_{t}$ being respectively even and odd in  $ 
\varepsilon_{n}$.  The $z$-axis is
perpendicular to the interfaces, and the CM region is delimited by $|z|
< d_{f}/2$, where $d_{f}$ is the thickness of the CM layer.  The
exchange field ${\bf J}$ is nonzero in the CM region, while the singlet
superconducting order parameter $\Delta_{s}$ is nonzero only in the S
regions. The S and CM parts can have different diffusion constants,
$D_s$ and $D_f$, and therefore also different superconducting
coherence lengths $\xi_{s,f}=\sqrt{D_{s,f}/2\pi T_{c}}$. For
simplicity, we assume that the two S regions, and also the two S/CM
interfaces, are characterized by identical parameters. Another
important length scale is the magnetic length
$\xi_J=\sqrt{D_f/J}$.

The exchange field ${\bf J}$
rotates within the $x-y$ plane in the CM film
with a spiral wave vector $Q{\bf e}_y$,
\begin{equation}
  {\bf J}(y)=J(\cos Qy,\sin Qy,0). \label{inhomJ}
\end{equation}
As a result
$f$ depends on both spatial coordinates $z$ and $y$.
The triplet vector ${\bf f}_{t}$ is found to be parallel to ${\bf J}$  
everywhere \cite{Cha2005}.
It is convenient to introduce chiral triplet components
$
f_{\pm}=\left( \mp f_{tx}+if_{ty}\right)
e^{\pm i Q y}. \label{conv}
$
In the CM layer, the singlet component $f_{s}$ and the two chiral
triplet components $f_{\pm}$ are then given by
\begin{equation}
f_{l}(z)= \sum_{\epsilon=\pm 1} \varphi_{l,\epsilon} \left[a_ 
{\epsilon}\cosh\left(k_{\epsilon} z \right)+b_{\epsilon}\sinh\left(k_ 
{\epsilon} z \right) \right], \label{sol}
\end{equation}
where $l=s$ or $\pm$, $\varphi_{s,\epsilon} = \eta_{\epsilon}$,
$\varphi_{-,\epsilon} = \epsilon$, $\varphi_{+,\epsilon} = -\epsilon$
and
\begin{eqnarray}
k_{\epsilon}  &=&\sqrt{2(\varepsilon_{n} + \epsilon i J \eta_{- 
\epsilon})/D_{f}}, \quad
\\
\eta_{\epsilon} &=& \left\{
\begin{array}{c}
\sqrt{1 -\eta^2}+ i \epsilon \eta \\
-i(\sqrt{\eta^2-1}- \epsilon \eta )\\
\end{array}
\quad \mbox{for} \quad
\begin{array}{c}
\eta \le 1\\
\eta > 1 \\
\end{array}
\right. ,
\end{eqnarray}
where $\eta = D_{f} Q^2/4J = (Q\xi_J)^2/4$. As the singlet component  
$f_{s}$, the chiral triplet components penetrate over the short  
length scale $\xi_{J}$ inside the chiral magnet.

The different coefficients $a_{\epsilon}$ and $b_{\epsilon}$ are  
determined
by boundary conditions
for the two S/CM interfaces (located at $z=\pm
d_{f}/2$). These
connect the $f$ on the
S side of the interface (denoted $z_{s}$) with the $f$
on the CM side at the interface (denoted $z_{f}$)
and read \cite{Kup1988}
\begin{eqnarray}
\gamma \xi_{f} \partial_z f_{l} (z_{f}) & = & \xi_{s} \partial_z f_ 
{l} (z_{s}), \label{B1} \\
\gamma_{b} \xi_{f}  \partial_z f_{l}(z_{f}) & = & \pm \left[f_{l}(z_ 
{s}) -f_{l}(z_{f}) \right],\label{B2}
\end{eqnarray}
for the triplet ($l=\pm$) and singlet ($l=s$) amplitudes. The
parameters $\gamma$ and $\gamma_{b}$ are related to the conductivity
mismatch between the two sides ($\gamma \xi_f/\xi_s=\sigma_f/\sigma_s 
$ with
the bulk conductivities $\sigma_{f}$ in CM and $\sigma_s$ in S)
and the boundary resistance, respectively.
The signs $\pm $ in Eq.~(\ref{B2}) refer to the interfaces at $z=\pm  
d_f/2$,
respectively.
In the following, we define the
short-hand notation $\delta^{(\pm )}=f_{s}(\pm d_{f}/2)$
for the singlet amplitudes at the interfaces.

Due to the leakage of pair correlations into the
central CM region, the amplitudes $\delta^{(\pm )}$ are expected to be
reduced compared with the bulk value in S. This inverse proximity effect
can be important in hybrid structures involving ferromagnets (see  
e.g. Ref. \onlinecite{Lof2007}). However,  the spatial dependences of  
$f_{s}$ as well as of the triplet components can be disregarded in S  
when
$\gamma \ll 1 + \gamma_{b} d_{f}/\xi_{f}$, and the rigid boundary  
conditions
hold (see, e.g., Ref. \cite{Gol}), with $\delta^{(\pm )} \approx \pi
\Delta_{s} e^{\pm i \phi/2}/
\sqrt{\varepsilon_{n}^{2}+\Delta_{s}^{2}}$, where $\phi$ is the phase
difference between the two superconductors.
Using Eqs. (\ref{B1})-(\ref{B2}) within this assumption, we express
$a_{\epsilon}$ and $b_{\epsilon}$ as functions of $\delta^{(\pm )}$
\begin{eqnarray}
a_{\epsilon }&=& \frac{\delta^{(+)}+\delta^{(-)}}{2} \;
\frac{1}{\left( \eta_{\epsilon }+ \eta_{-\epsilon } \right){\cal A}_ 
{\epsilon }} , \label{D1}\\
b_{\epsilon }&=&\frac{\delta^{(+)}-\delta^{(-)}}{2} \; \frac{1}{\left 
(\eta_{\epsilon } + \eta_{-\epsilon } \right){\cal B}_{\epsilon } },  
\label{D2}
\end{eqnarray}
where
${\cal A}_{\epsilon }= \cosh (x_{\epsilon })+ \gamma_{b} k_ 
{\epsilon } \xi_{f} \sinh (x_{\epsilon })$,
${\cal B}_{\epsilon }= \sinh (x_{\epsilon }) + \gamma_{b} k_ 
{\epsilon } \xi_{f} \cosh (x_{\epsilon })$,
and $x_{\epsilon }=k_{\epsilon }d_f/2$.

The current flowing through the S-CM-S junction is
\begin{equation}
I= 2e \frac{D_{f}}{\pi} N_{f} {\cal S} T \sum_{n} \mathrm{Im} \left[
f_{s}^{\ast} \partial_{z} f_{s}-f_{tx}^{\ast} \partial_{z} f_{tx}-f_ 
{ty}^{\ast} \partial_{z} f_{ty}
\right], \label{C1}
\end{equation}
where $N_{f}$ is the Fermi-level density of states per spin in CM and ${\cal S}$ is
the cross-section area. We insert 
$f_{tx}^{\ast} \partial_{z} f_{tx}+f_{ty}^{\ast} \partial_{z} f_{ty}= 
\left(
f_{-}^{\ast} \partial_{z}f_{-}+f_{+}^{\ast} \partial_{z} f_{+}
\right)/2 $ and
Eq. (\ref{sol}) in Eq. (\ref{C1}), and
express $I$ as a function of $a_{\epsilon}$ and
$b_{\epsilon}$ as
\begin{equation}
I =4 e \frac{D_{f}}{\pi} N_{f} {\cal S} T  \sum_{n \geq 0}\sum_ 
{\epsilon,\epsilon'} \mathrm{Im}
\left[
(\eta_{\epsilon'}^\ast \eta_{\epsilon}-\epsilon' \epsilon )
a_{\epsilon'}^{\ast} b_{\epsilon}
k_{\epsilon }
\right] . \label{C3}
\end{equation}
For $\eta < 1$ only the terms with $\epsilon \ne \epsilon' $  
contribute, while for
$\eta > 1$ only the terms with $\epsilon =\epsilon' $ contribute [the
case $\eta=1$ is defined via the corresponding limit in Eq.~(\ref{C3})].
In agreement with current conservation, the dependence on $z$ vanishes.

\begin{figure}[t]
\includegraphics[width=0.9\columnwidth]{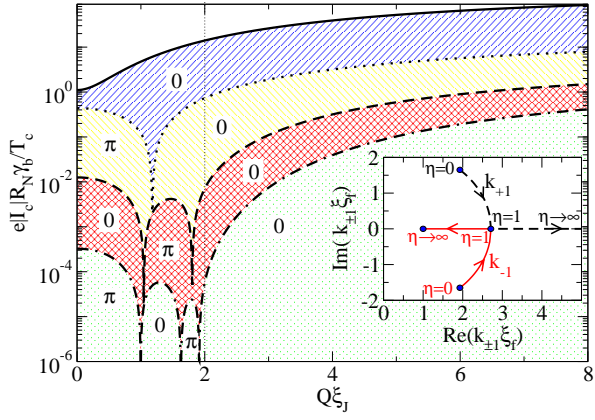}
\caption{(Color online) Josephson critical current $I_{c}$ versus the
   spiral wave vector $Q$
   for a few thicknesses of the ferromagnet:
   curves from top to bottom
   $d_f/\xi_f=0.1$, $1$, $3$, $5$.
   Here $T=0.1T_c$ and $J=20 T_{c}$.
   The inset shows the flow of the real and imaginary parts of the
   eigenvalues $k_{\pm 1}$ with varying $\eta=(Q\xi_J)^2/4$ for
   $\varepsilon_n=\pi T_c$.}
\label{Fig2}
\end{figure}

It then follows from Eqs. (\ref{D1}), (\ref{D2}), and (\ref{C3}) that  
the
current-phase relation reduces to sinusoidal form $I=I_{c} \sin (\phi)$.
Close to $T_{c}$
we get for $I_{c}$
\begin{equation}
I_{c}R_N= 4 V_{0} \left(\frac{d_{f}}{\xi_{f}}+2 \gamma_{b} \right)
  \sum_{n \geq 0}\sum_{\epsilon=\pm 1} \, \frac{T_{c}^{2}} 
{\varepsilon_{n}^{2}} \;
  \frac{k_{\epsilon} \xi_{f}\eta_{\epsilon} }{{\cal A}_{\epsilon} 
{\cal B}_{\epsilon} (\eta_{\epsilon}+\eta_{-\epsilon})}
,
\label{R1}
\end{equation}
where
$R_{N}=(d_{f}+2 \gamma_{b}\xi_{f})/\sigma_{f}{\cal S}$ is the
normal state resistance, $\sigma_{f}=2 e^{2}N_{f}D_{f}$ is the
conductivity of the CM layer, and $V_{0}=\pi \Delta_{s}^{2}/4 e T_{c}$.
On the other hand,
for low barrier transparencies ($\gamma_{b} \gg 1$) \cite{valid}, we
have ${\cal A}_{\epsilon } \approx \gamma_{b}
k_{\epsilon } \xi_{f} \sinh(k_{\epsilon  } d_{f}/2)$ and $ {\cal B}_ 
{\epsilon } \approx \gamma_{b} k_{\epsilon } \xi_{f}
\cosh(k_{\epsilon } d_{f}/2)$, which
lead to
\begin{equation}
I_{c}R_N=  \frac{4 \pi}{\gamma_{b}} \frac{T}{e}
  \sum_{n \geq 0}\sum_{\epsilon=\pm 1} \, \frac{\Delta_{s}^{2}} 
{\varepsilon_{n}^{2}+\Delta_{s}^{2}} \; \;
  \frac{\eta_{\epsilon}/(\eta_{\epsilon}+\eta_{-\epsilon}) }{k_ 
{\epsilon} \xi_{f} \sinh(k_{\epsilon}d_{f})}
.
  \label{R2}
\end{equation}
In the absence of inhomogeneity ($Q=0$), we then recover expressions for
the critical current in the literature \cite{Gol}.
Note that the temperature $T$ appears through several terms in
Eq. (\ref{R2}), such
as $\Delta_{s}$ (here we assume the BCS temperature dependence),
$\varepsilon_{n}$ and $k_{\epsilon}$.

\begin{figure}[t]
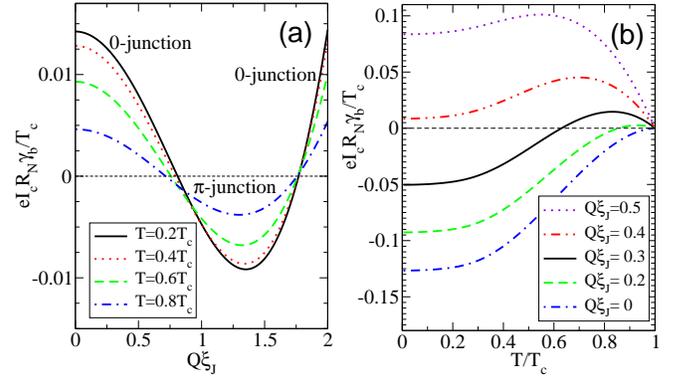

\includegraphics[width=0.49\columnwidth]{IcVsQ3.eps}
\includegraphics[width=0.49\columnwidth]{IcVsT3.eps}
\caption{(Color online) (a) Critical current versus spiral wave vector
   for a few temperatures, in the region $0\leq Q\xi_J\leq 2$ where
   $0-\pi$ transitions are possible.
   Here $d_f=2.7 \xi_f$ and $J=20T_c$.
   (b) The $0-\pi$ transition is
   observable as function of temperature for a certain thickness (here
   $d_f=0.45\xi_f$)
   by tuning the spiral wave vector. The other model
   parameters are the same as in (a).}
\label{Fig3}
\end{figure}

In the following we study the influence of an exchange field with
chiral order on the Josephson effect on the basis of
Eq. (\ref{R2}). For small thicknesses $d_{f}$ we have used the more general expression (\ref{C3}) to verify that Eq.  (\ref{R2}) indeed is applicable in the parameter range we consider.
  As we show in
Fig. \ref{Fig2}, the chiral magnetic order introduces
a surprizingly rich behavior:
the magnitude of $I_{c}$
as function of increasing wave vector $Q$
presents initial oscillations
and suppression, followed by increase and final saturation. Depending
on the thickness of the CM layer, there can be one or
several $0-\pi$ and $\pi-0$ transitions as function of the spiral
order wave vector $Q$. Above a certain value of $Q$ ($Q\xi_J=2$
indicated by the vertical line in the figure) $I_{c}$ is positive
independently of other model parameters, meaning that the junction
phase difference is stabilized at zero. Physically, this can be
understood as an averaging out of the exchange field within one
magnetic length $\xi_J$. Technically, this critical value of $Q$
separates a region with complex eigenvalues $k_{\epsilon}$ ($\eta<1$,
oscillating $I_c$) from a region with real $k_{\epsilon}$ ($\eta>1$,
monotonously increasing $I_c$), see the inset of Fig. \ref{Fig2}. For  
$\eta<1$, the
complex
$k_{\epsilon}$ leads to a non-monotonic
dependence of $I_c$ as function of Q. In the large-$Q$ limit, the
Josephson critical current for a junction with
a normal metal is recovered.

In Fig. \ref{Fig3}(a) we study in more detail the critical current
within the region $0\leq Q\xi_J\leq 2$ supporting oscillations. For an
intermediately thick magnetic film (here $d_f=2.7\xi_f$) it is
possible to see both $0-\pi$ and $\pi-0$ transitions as function of
$Q$, with a reasonably large critical current. The phase transitions
shift to lower values of $Q$ with increasing temperature.
As seen in Fig. \ref{Fig3}(b), the spiral order can also induce
$0-\pi$ transitions as function of temperature for certain parameter  
ranges.

\begin{figure}[t]
\includegraphics[width=0.50\columnwidth]{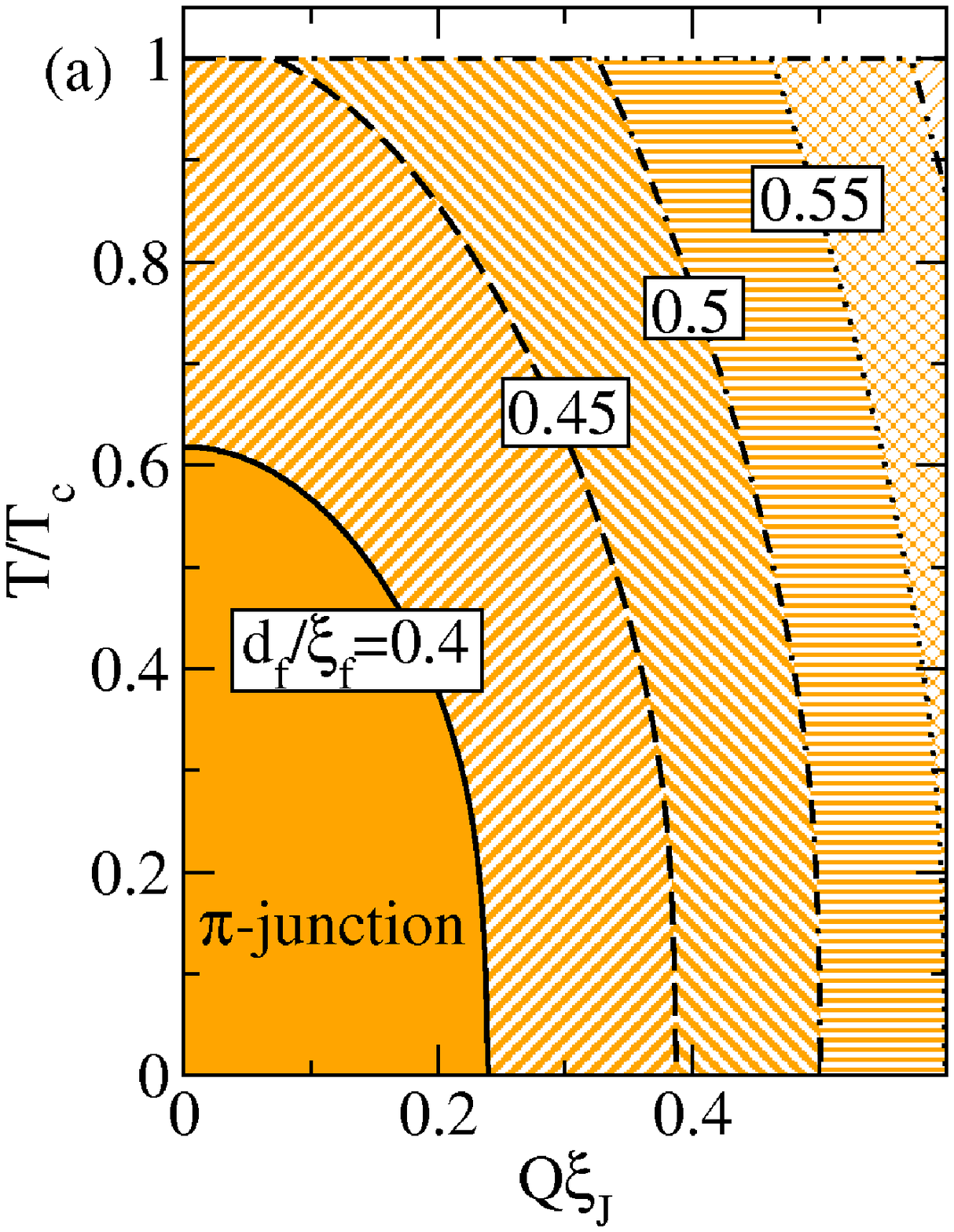}
\includegraphics[width=0.48\columnwidth]{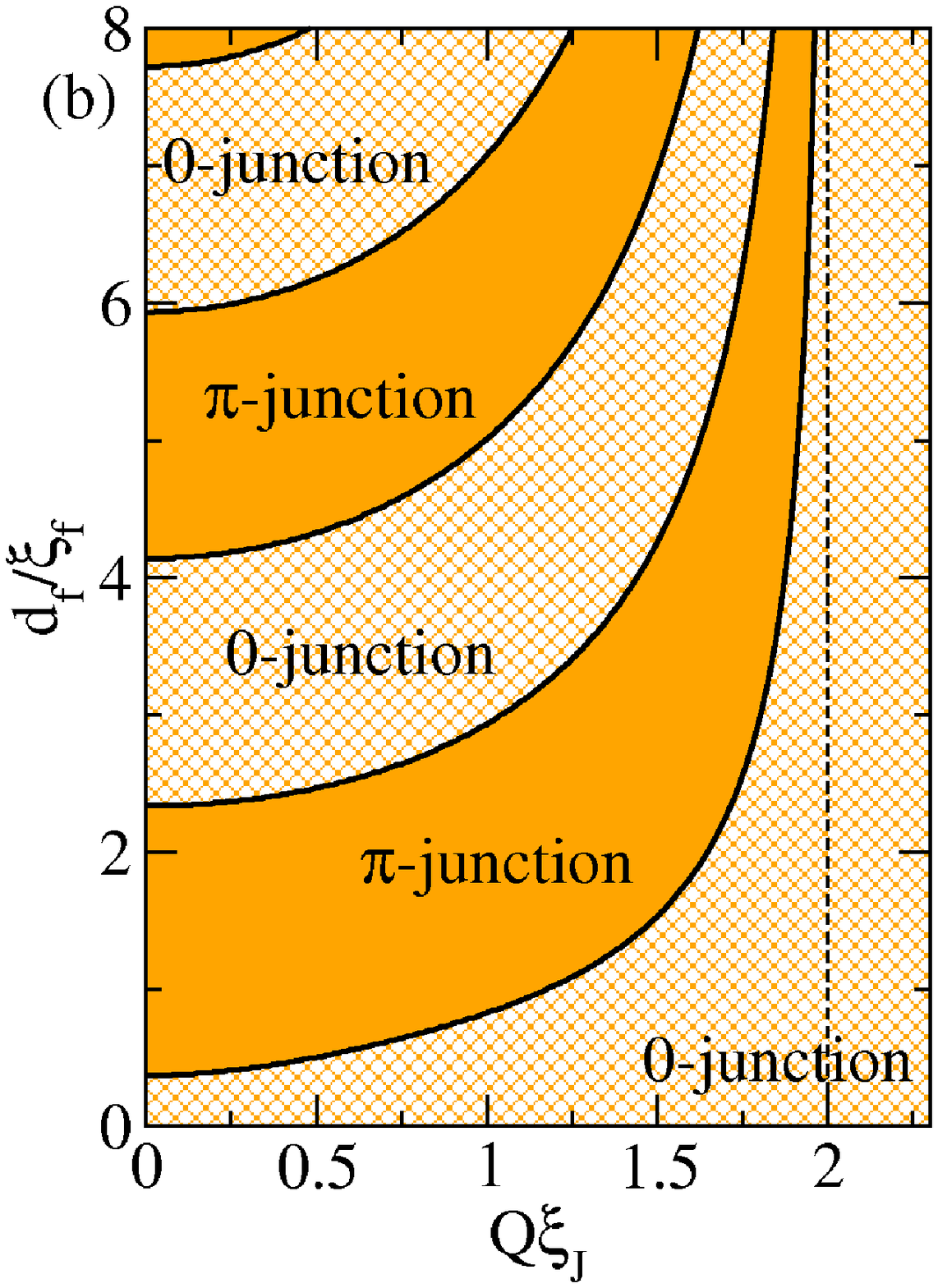}
\caption{(Color online) (a): ($T$-$Q$) phase diagram for a Josephson
   junction with a chiral magnet ($J=20T_c$) between two singlet
   superconductors.  The transition from a $\pi $-junction at smaller
   chiral wave vector $Q$ to a $0$-junction at larger $Q$ is indicated
   for several thicknesses of the ferromagnetic layer.  (b):
   corresponding low-T ($d_f$-$Q$) phase diagram
   ($T=0.1T_c$).  }
\label{Fig4}
\end{figure}

Phase-diagrams of the $\pi-0$ transitions are presented in
Fig.~\ref{Fig4}. We see that in the low-$T$ region [panel (a)] the
phase transition line $T_{\pi-0}(Q)$ develops a very steep slope.
This insensitivity to temperature variations can be of importance for
device applications. Although at ultra-low temperatures a more
sophisticated theory than the mean field approach presented here
should be used, our results in Fig.~\ref{Fig4} give a strong
indication of a $\pi-0$ transition as a function of $Q$ also at zero
temperature.  Thus, the
system of a chiral magnet sandwiched between two superconductors is
of potential interest for the study of critical behavior near a
quantum critical point.

In the right panel of Fig. \ref{Fig4} the very different behaviors for
$Q\xi_J<2$ and $>2$ are also seen. For $Q\xi_J<2$ the spiral order
shifts the transition lines towards thicker magnetic
films, but the transition line never disappears from the phase
diagram. Only in the region $Q\xi_J>2$ is the averaging of the
exchange field over the magnetic length so effective as to prevent
$0-\pi$ transitions.

In summary, we have studied the Josephson effect in an S-CM-S junction
in the presence of an in-plane cycloidal spin spiral structure in   
the magnet.  We have found that the presence of a spin spiral can
change the ground state of the Josephson junction, and lead to a
transition between a $\pi$-junction and a $0$-junction for a critical
spiral wave vector. The dependences of the Josephson effect on magnet
thickness and on temperature depend sensitively on the wave vector of
the chiral order in the magnet. We predict that a
quantum-critical point should exist in the phase-diagram for suitably
chosen sample parameters. We expect that these effects will have
potential applications for new types of functional nanoscale
structures.
The ultimate goal for the future is to tune Josephson junctions
with one or more chiral magnets by controlling
the phases or magnitudes of the spiral magnetic wave vectors.

We would like to thank Gerd Sch\"on for valuable contributions to  
this work.
T.L. acknowledges support from the Alexander von Humboldt Foundation.

{\em Note added}. - After submission, we became aware of work by
Crouzy {\em et al.} \cite{Crouzy2007}, who study in-plane magnetic
N\'eel domain walls. Their model is markedly different from ours, but
leads to similar findings about the periodicity of 0 to $\pi$
transitions with the magnetic inhomogeneity.


\begin{thebibliography}{99}
\bibitem{valid}
We assume that the following inequalities hold $\left(\gamma_{b} k_{\epsilon} \xi_{f} \right)^{-1}
\ll \tanh(x_{\epsilon}) \ll \gamma_{b} k_{\epsilon} \xi_{f} 
$.
\end{thebibliography}
\end{document}